\documentclass[epj]{svjour}
\usepackage{graphicx}

\begin{document}

\title{Conductance quantization in graphene nanoribbons: Adiabatic approximation}
\titlerunning{Conductance quantization in graphene nanoribbons}
\author{M. I. Katsnelson}
\institute{Institute for Molecules and Materials, Radboud
University Nijmegen, 6525 ED Nijmegen, The Netherlands}
\date{Received: date / Revised version: date}

\abstract{A theory of electron states for graphene nanoribbons
with a smoothly varying width is developed. It is demonstrated
that the standard adiabatic approximation allowing to neglect the
mixing of different standing waves is more restrictive for the
massless Dirac fermions in graphene than for the conventional
electron gas. For the case of zigzag boundary conditions, one can
expect a well-pronounced conductance quantization only for highly
excited states. This difference is related to the relativistic
Zitterbewegung effect in graphene.
}
\maketitle

The experimental discovery of a truly two-dimensional allotrope of carbon,
graphene~\cite{kostya0,kostya1} and of the massless Dirac character of its
electronic energy spectrum~\cite{kostya2,kim} has initiated an enormously
growing interest in this field (for review, see Refs.~\cite
{reviewGK,reviewktsn}). One of the most exciting aspects of the problem is
the hope to develop novel carbon-based electronics. Very recently, the
experimental realization of quantum dots~\cite{reviewGK} and carbon
nanoribbons~\cite{chen,kim2} has been announced, the former demonstrating
single-electron transistor (SET) effect~\cite{reviewGK}.

The conductance quantization in the ballistic regime~\cite
{been,glazman,payne,imry,kaveh} is one of the most important physical
phenomena determining the functioning of such nanodevices. It was considered
recently~for the case of ideal graphene stripe \cite{castroneto} and for the
case of confinement due to a smooth external electrostatic potential \cite
{efetov}. The experimental situation~\cite{reviewGK,chen,kim2} corresponds
rather to the case of electron confinement due to a curvilinear shape of the
graphene samples than to an external field. The description of the
penetration of electron waves through constrictions in the nanoribbons
requires a different theoretical approach. Numerical calculations of
electronic transport in graphene nanoribbons demonstrating a very
interesting ``valley filter'' effect have been recently published~\cite
{valley}. However, a general theoretical analysis of the situation is still
absent. Here we present an analytical theory of conductance quantization in
graphene nanoribbons based on the adiabatic approximation~\cite{glazman,imry}%
. The latter means a separation of the electron motion in the
directions perpendicular and along the stripe. For nonrelativistic
electrons the adiabatic approximation requires only the smoothness
of the shape of the stripe boundary and results in the
quantization of the conductance. For graphene, the situation turns
out to be essentially dependent on the boundary conditions. It
will be shown that for the zigzag boundaries this theory is
essentially different from that for nonrelativistic electrons and,
in general, there is no reason to expect an adiabatic regime and
well-pronounced conductance jumps for the lowest states of the
ribbon.

The two-component wave function $(u,v)$ for charge carriers in graphene with
wave vectors close to the $K$ point is described by the Dirac equation
\begin{eqnarray}
\left( \frac \partial {\partial x}+i\frac \partial {\partial y}\right) u
&=&ikv,  \nonumber \\
\left( \frac \partial {\partial x}-i\frac \partial {\partial y}\right) v
&=&iku  \label{dirac}
\end{eqnarray}
where $k=E/\hbar v_F,$ $E$ is the electron energy and $v_F\simeq 10^6$ m/s
is the Fermi velocity~\cite{kostya2,kim}; for the other valley $K^{\prime }$
the signs before $\partial /\partial y$ are opposite. Let us consider first
the case of a uniform graphene strip of width $L$ along the $y$-axis, $%
\left| y\right| <L/2.$ To specify the problem one has to choose boundary
conditions at the edges~\cite{brey}.

We start with the case of zigzag edges where $u\left( y=-L/2\right) $ $%
=0,v\left( y=L/2\right) $ $=0.$ The energy spectrum is discrete, $E_j=\hbar
v_Fk_j$ where
\begin{equation}
k_j=\frac{\pi j}L,
\begin{array}{cc}
& j=\pm \frac 12,\pm \frac 32,...
\end{array}
\label{kj}
\end{equation}
and the wave functions have the form
\begin{eqnarray}
u_j\left( y\right) &=&\frac 1{\sqrt{L}}\cos k_j\left( y-L/2\right) ,
\nonumber \\
v_j\left( y\right) &=&-\frac 1{\sqrt{L}}\sin k_j\left( y-L/2\right) .
\label{uv}
\end{eqnarray}

Consider now the case of a smoothly varying strip width, $L\rightarrow
L\left( x\right) ,\left| dL/dx\right| $ $\ll 1.$ Following a general scheme~%
\cite{glazman,imry} one can try a solution of equation (\ref{dirac}) as an
expansion
\begin{eqnarray}
u\left( x,y\right)  &=&\sum\limits_jc_j\left( x\right) u_j^{(x)}\left(
y\right) ,  \nonumber \\
v\left( x,y\right)  &=&\sum\limits_jc_j\left( x\right) v_j^{(x)}\left(
y\right)   \label{expan}
\end{eqnarray}
where $u^{(x)},v^{(x)}$ are the functions (\ref{uv}) with the replacement $%
L\rightarrow L\left( x\right) .$ The functions (\ref{expan}) satisfy by
construction the boundary conditions. By substituting the expansion (\ref
{expan}) into Eq.(\ref{dirac}), multiplying the first equation by $%
\left\langle v_j\right| $ and the second one by $\left\langle u_j\right| $
one finds:
\begin{eqnarray}
\sum\limits_{j^{\prime }}\left[ \frac{dc_{j^{\prime }}}{dx}\left\langle
v_j|v_{j^{\prime }}\right\rangle +c_{j^{\prime }}\left\langle v_j|\frac{%
dv_{j^{\prime }}}{dx}\right\rangle \right]  &=&i\sum\limits_{j^{\prime
}}\left( k-k_{j^{\prime }}\right)   \nonumber   \\
&&c_{j^{\prime }}\left\langle v_j|u_{j^{\prime }}\right\rangle , \nonumber \\
\sum\limits_{j^{\prime }}\left[ \frac{dc_{j^{\prime }}}{dx}\left\langle
u_j|u_{j^{\prime }}\right\rangle +c_{j^{\prime }}\left\langle u_j|\frac{%
du_{j^{\prime }}}{dx}\right\rangle \right]  &=&i\sum\limits_{j^{\prime
}}\left( k-k_{j^{\prime }}\right)   \nonumber  \\
&&c_{j^{\prime }}\left\langle u_j|v_{j^{\prime }}\right\rangle .
\label{exact}
\end{eqnarray}

This equation is formally exact. As a first step to the adiabatic
approximation, one should neglect the terms with $\left\langle v_{j}|\frac{%
dv_{j^{\prime }}}{dx}\right\rangle $ and $\left\langle u_{j}|\frac{%
du_{j^{\prime }}}{dx}\right\rangle $ which is justified by the smallness of $%
dL/dx$, as in the case of nonrelativistic electrons~\cite{glazman,imry}.

To proceed further we need to calculate the overlap integrals $\left\langle
\phi _1|\phi _2\right\rangle =\int\limits_{-L/2}^{L/2}dy\phi _1^{*}\phi _2$
for different basis functions:
\begin{eqnarray}
\left\langle u_j|u_{j^{\prime }}\right\rangle &=&\frac 12\left( \delta
_{jj^{\prime }}+\delta _{j,-j^{\prime }}\right) ,  \nonumber \\
\left\langle v_j|v_{j^{\prime }}\right\rangle &=&\frac 12\left( \delta
_{jj^{\prime }}-\delta _{j,-j^{\prime }}\right) ,  \nonumber \\
\left\langle u_j|v_{j^{\prime }}\right\rangle &=&\left\langle v_{j^{\prime
}}|u_j\right\rangle =\left\{
\begin{array}{ccc}
-\frac 1{\pi \left( j^{\prime }-j\right) }, &  & j^{\prime }-j=2n+1, \\
-\frac 1{\pi \left( j^{\prime }+j\right) }, &  & j^{\prime }-j=2n,
\end{array}
\right.  \label{overlap}
\end{eqnarray}
where $n$ is integer. Substituting Eq.(\ref{overlap}) into Eq.(\ref{exact})
and neglecting the nonadiabatic terms with the matrix elements of the
operator $d/dx$, we obtain after simple transformations:
\begin{equation}
\frac{dc_j\left( x\right) }{dx}=-\frac{2i}\pi \sum\nolimits_{j^{\prime
}}^{\prime }\frac{\left[ k-k_{j^{\prime }}\left( x\right) \right] }{%
j+j^{\prime }}c_{j^{\prime }}\left( x\right)  \label{c}
\end{equation}
where the sum is over all $j^{\prime }$ such that $j^{\prime }-j$ is even.

Until now we did transformations and approximations which are identical to
those used in the case of nonrelativistic electrons. However, we still have
a coupling between different standing waves so we still cannot prove that
the electron transmission through the constriction is adiabatic. To prove
the latter we need one more step, namely, a transition from the discrete
variable $j$ to real one and a replacement of the sums by integrals in the
right-hand-side of Eq.(\ref{c}): $\sum\nolimits_j^{\prime }...\rightarrow
\frac 12\mathcal{P}\int dj...$ where $\mathcal{P}$ is the symbol of
principle value. This step is justified by assuming that $kL\gg 1,$ i.e.,
only for highly excited states. For the low-lying electron standing waves it
is difficult to see any way to simplify essentially the set of equations (%
\ref{c}) for the coupled states.

For any function $f\left( z\right) $ analytical in the upper (lower) complex
half-plane one has
\begin{equation}
\int\limits_{-\infty }^\infty dxf\left( x\right) \frac 1{x-x_1\pm i0}=0
\label{identity}
\end{equation}
\\\\or, equivalently,
\begin{equation}
\int\limits_{-\infty }^\infty dxf\left( x\right) \frac{\mathcal{P}}{x-x_1}%
=\pm i\pi f\left( x_1\right) .  \label{iden1}
\end{equation}
Assuming that $c_j\left( x\right) $ is analytical in the lower halfplane as
a function of complex variable $j$ one obtains, instead of Eq.(\ref{c})
\begin{equation}
\frac{dc_j\left( x\right) }{dx}=\left[ k+k_j\left( x\right) \right]
c_{-j}\left( x\right) .  \label{c1}
\end{equation}
Similar, taking into account that $c_{-j}\left( x\right) $ is analytical in
the \textit{upper} halfplane as a function of complex variable $j$ we have
\begin{equation}
\frac{dc_{-j}\left( x\right) }{dx}=\left[ k_j\left( x\right) -k\right]
c_j\left( x\right) .  \label{c2}
\end{equation}
At last, differentiating Eq.(\ref{c1}) with respect to $x$, neglecting the
derivatives of $k_j\left( x\right) $ due to the smallness of $dL/dx$ and
taking into account Eq.(\ref{c2}) we find
\begin{equation}
\frac{d^2c_j\left( x\right) }{dx^2}+\left[ k^2-k_j^2\left( x\right) \right]
c_j\left( x\right) =0.  \label{final}
\end{equation}

Further analysis completely follows that for the nonrelativistic case~\cite
{glazman} where $k^{2}$ and $k_{j}^{2}\left( x\right) $ play the roles of
energy and potential, respectively. The potential is quasiclassical for the
case of smoothly varying $L(x)$. Therefore, the transmission coefficient is
very close to one if the electron energy exceeds the energy of the $j$th
level in the narrowest place of the constriction, and exponentially small,
otherwise. Standard arguments based on the Landauer formula~\cite
{been,glazman,payne,imry,kaveh} prove the conductance quantization in this
situation.

At the same time, for the \textit{lowest} energy levels the replacement of
sums by integrals in Eq.(\ref{c1}) cannot be justified and thus the states
with different $j$'s are in general coupled even for a smooth constriction ($%
\left| dL/dx\right| \ll 1$). Therefore electron motion along the stripe is
strongly coupled with that in the perpendicular direction and different
electron standing waves are essentially entangled. In this situation there
is no general reason to expect sharp jumps and well-defined plateaus in the
energy dependence of the conductance. This means that the criterion of
adiabatic approximation is more restrictive for the case of Dirac electrons
than for the nonrelativistic ones. The formal reason is an overlap between
components of the wave functions with different pseudospins or,
equivalently, between hole component of the state $j$ with the electron
component of the state $j^{\prime }\neq j$. This coupling is a reminiscence
of the Zitterbewegung of Dirac electrons determining the finite conductivity
and anomalous shot noise in graphene in the limit of small charge carrier
concentration~\cite{zitter,noise}. Effectively, it works as a kind of
intrinsic ``disorder'' and therefore it is not surprising that it destroys
the ballistic regime near the Dirac point. Interestingly, the kinetic
equation that takes into account the Zitterbewegung effects also contains
some ``$\mathcal{P}$-terms'' which are absent in the standard Boltzmann
equation; these terms become irrelevant for large enough Fermi energy~\cite
{aus}.

Consider now the case of armchair edges. The boundary conditions are coupled
the components of Dirac spinors at $K$ valley $u,v$ with those at $K^{\prime
}$ valley $\overline{u},\overline{v}:$%
\begin{eqnarray}
u\left( -L/2\right) &=&\overline{u}\left( -L/2\right) ,  \nonumber \\
v\left( -L/2\right) &=&\overline{v}\left( -L/2\right) ,  \nonumber \\
u\left( L/2\right) &=&e^{2\pi i\nu }\overline{u}\left( L/2\right) ,
\nonumber \\
v\left( L/2\right) &=&e^{2\pi i\nu }\overline{v}\left( L/2\right) ,
\label{arm}
\end{eqnarray}
where $\nu =0,\pm 2/3$, depending on the number of rows in the strip \cite
{brey}. The eigenmodes in that case are just plane waves \cite{brey}
\begin{eqnarray}
u_j\left( y\right) &=&-iv_j\left( y\right) =\frac 1{\sqrt{2L}}\exp \left(
ik_jy\right) ,  \nonumber \\
\overline{u}_j\left( y\right) &=&-i\overline{v}_j\left( y\right) =\frac 1{%
\sqrt{2L}}\exp \left( -ik_jy\right) ,  \label{armuv}
\end{eqnarray}
\begin{equation}
k_j=\left( j+\nu \right) \pi /L,
\begin{array}{cc}
& j=0,\pm 1,\pm 2,...
\end{array}
\label{wave}
\end{equation}
A general solution satisfying the boundary conditions (\ref{arm}) can be
probed as
\begin{eqnarray}
u\left( x,y\right) &=&\sum\limits_jc_j\left( x\right) \exp \left[ ik_j\left(
x\right) y\right] ,  \nonumber \\
v\left( x,y\right) &=&i\sum\limits_jb_j\left( x\right) \exp \left[
ik_j\left( x\right) y\right] ,  \nonumber \\
\overline{u}\left( x,y\right) &=&\sum\limits_jc_j\left( x\right) \exp \left[
ik_j\left( x\right) \left( L-y \right) \right] ,  \nonumber \\
\overline{v}\left( x,y\right) &=&i\sum\limits_jb_j\left( x\right)
\exp \left[ ik_j\left( x\right)\left( L-y \right) \right] .
\label{arm11}
\end{eqnarray}
Substituting this into the Dirac equation (\ref{dirac}) one obtains
\begin{eqnarray}  \label{arm12}
\sum\limits_j\exp \left[ ik_j\left( x\right) y\right] \left\{ \frac{dc_j}{dx}%
+\left( kb_j-k_jc_j\right) +i\frac{dk_j}{dx}yc_j\right\} &=&0,  \nonumber \\
\sum\limits_j\exp \left[ ik_j\left( x\right) y\right] \left\{ \frac{db_j}{dx}%
+\left( k_jb_j-kc_j\right) +i\frac{dk_j}{dx}yb_j\right\} &=&0. \nonumber \\
\end{eqnarray}

Let us neglect first the nonadiabatic terms proportional to $\frac{dk_j}{dx}$
in these equations. They will be satisfied for sure if all coefficients at
the plane waves vanish, which is equivalent to the set of equations
\begin{eqnarray}
\frac{d\left( c_j+b_j\right) }{dx}+\left( k+k_j\right) \left( b_j-c_j\right)
&=&0,  \nonumber \\
\frac{d\left( b_j-c_j\right) }{dx}+\left( k_j-k\right) \left( c_j+b_j\right)
&=&0.  \label{arm13}
\end{eqnarray}
Differentiating them with respect to $x$ and neglecting, again, the
derivatives of $k_j$ we find the effective Schr\"odinger equation (\ref
{final}) and the same equation for $b_j$. Thus, in contrast with the case of
zigzag edges, for the armchair edges a standard picture of conductance
quantization should be valid for all states, similar to nonrelativistic
electron gas.

However, there is another problem which makes the adiabatic
approximation for this case problematic. The wave numbers
(\ref{wave}) can depend on $x$ not only due to the stripe length
but also due to different number of rows in the stripe which makes
$d\nu /dx$ a source of sharp random potential. It is very
difficult to investigate this effect analytically in the framework
of the approach under consideration. It was argued recently based
on numerical results and qualitative considerations that this kind
of randomness should be of crucial importance for the graphene
nanoribbons with the armchair edges \cite{martin}.

It would be very interesting to check experimentally the possible difference
in the conductance behavior for the nanoribbons with zigzag and armchair
edges. For the former case, the theory predicts essential difference of
behavior at the crossing of low-lying and highly excited energy levels in
the quantum point contact situation, that is, for a narrow constriction of
the graphene nanoribbons.

\textit{Acknowledgements}. I am thankful to Andre Geim, Kostya Novoselov,
and Annalisa Fasolino for helpful discussions. This work was supported by
the Stichting voor Fundamenteel Onderzoek der Materie (FOM), the Netherlands.

\end{document}